\documentclass[preprint,12pt]{elsarticle}
\usepackage{etoolbox}
\patchcmd{\MaketitleBox}{\footnotesize\itshape\elsaddress\par\vskip36pt}{\footnotesize\itshape\elsaddress\par\parbox[b][36pt]{\linewidth}{\vfill\hfill\textnormal{August 30, 2016}\hfill\null\vfill}}{}{}%
\patchcmd{\pprintMaketitle}{\footnotesize\itshape\elsaddress\par\vskip36pt}{\footnotesize\itshape\elsaddress\par\parbox[b][36pt]{\linewidth}{\vfill\hfill\textnormal{August 30, 2016}\hfill\null\vfill}}{}{}%

\makeatletter

\def\ps@pprintTitle{%
 \let\@oddhead\@empty
 \let\@evenhead\@empty
 \def\@oddfoot{}%
 \let\@evenfoot\@oddfoot}

\makeatother

\usepackage{BP}
\usepackage[english]{babel}
\usepackage[utf8x]{inputenc}
\usepackage{amsmath}
\usepackage{mathrsfs}
\usepackage{amssymb}
\usepackage{graphicx}

\begin{document}
\begin{frontmatter}
\title{Maximal Sections of Sheaves of Data over an Abstract Simplicial Complex}
\author{Brenda Praggastis}

\begin{abstract}
We employ techniques from topological data analysis to model sensor networks. Our approach to sensor integration uses the topological method of sheaves over cell complexes. The internal consistency of data from individual sensors is determined by a set of consistency functions assigned to elements of the complex. Using these functions we determine, for any collection of data, the unique set of maximal sections of consistent data received from the sensors.  We offer a proof for the existence and uniqueness of these sections and illustrate the ideas with examples.
\end{abstract}
\end{frontmatter}

The association of data with a sensor network lends itself naturally
to the question of consistency. Data is received from various sensors, some of which should agree but may not. How does one decide which sensors are providing accurate information? Which sensors are providing data which are consistent with each other? How many possible consistent pictures can one obtain from a network? 

Sheaf theory provides a mathematical framework for the representation and analysis of interrelationships between sensors in a network \cite{Robinson3}\cite{Robinson1}. An extensive discussion of sheaves and their applications may be found in \cite{Curry}.   By modeling the network as a cell complex covered by a sheaf of data we describe what is meant by consistency and algorithmically compute the maximal sections over the network.

First we outline the bare essentials of a cell complex $\X$ which models the network. Let $\D =\{x_{\alpha}\}$ be the set of variables returned from a sensor network. Let $\V = \{v_i\}$ be the finite set of sensors, each returning values for some subset $\D(v_i) \subset \D$. For example a weather station could be a sensor $v$ which returns temperature $x_1$, atmospheric pressure $x_2$, humidity $x_3$, and windspeed $x_4$; in this case $\D(v) = \{x_i\}_{i=1}^4$.
We let $\V$ be the $0$-cells or vertex set for $\X$. 
The subsets formed by two vertices $\{v_1, v_2\}$ such that $\D(v_1) \cap \D(v_2) \neq \emptyset$ are the $1$-cells or edges in $\X$. Similarly, any subset of $k+1$ vertices $\sigma = \{v_0, v_1, ..., v_k\}$ such that $\cap \D(v_i) \neq \emptyset$ is a $k$-cell in $\X$. For any cell $\sigma$ in $\X$, define $\D(\sigma) = \cap_{v \in \sigma} \D(v)$, that is the set of variables shared by the vertices which formed the cell.\footnote{This is just one way to form the $k$-cells. Hyperedges in the cell complex could depend on some other relationship between the variables returned by the sensors.}
 
The collection of cells, $\X$, is an abstract simplicial cell complex. For if $\sigma$ is a $k$-cell in $\X$ generated by a set of vertices $\V_i \subset \V$ then $\cap_{v \in \V_i}\D(v) \neq \emptyset$ and the same must be true for any $\V_j \subset \V_i$. So $\X$ is closed under taking subsets. For each $\sigma \in \X$ define
$\str{\sigma} = \{ \tau \in \X : \sigma \subset  \tau  \}$. 
We give $\X$ the Alexandrov topology generated by $\{\str{v}\}_{v\in \V}$. For each $\sigma \in \X$, $\str{\sigma}$ is the smallest open set in $\X$ containing $\sigma$.

We now define a sheaf of sets $\shf$ over $\X$. 
For each open set $U \subset \X$ let $\D(U) = \cup_{\sigma \in U} \D(\sigma)$. 
Let $\shm{U}$ be the set of all possible assignments of values to the variables in $\D(U)$.
Define a section $s = \{s(x)\}_{x \in \D(U)} \in \shm{U}$ to be a single assignment of data values to each variable in $\D(U)$. Then $\shm{U}$ is the set of sections over $U$. A global section is a single assignment of values to each variable in $\D$. Note if $U'$ is an open subset of $U$ then $\D(U') \subset \D(U)$. Hence we may define a restriction map $\shr{U' \subset U}:\shm{U} \rightarrow \shm{U'}$ such that if $s \in \shm{U}$ then $\shr{U' \subset U}(s)(x) = s(x)$ for each $x \in \D(U')$, and we write $s|_{U'}$ to reference $s$ restricted to the variables in $\D(U')$. The locality and gluing properties of $\shf$ can be easily checked as two sections are equal on an open set $U$ if and only if they assign the same value to each variable in $\D(U)$.\footnote{Here we use projection maps for simplicity of demonstration only. Any restriction map which satisfies the locality and gluing properties of a sheaf may be used.}

Suppose we have a set of signals coming from the sensors in a network. In our model this corresponds to an assignment of values to the variables associated with that sensor.  For each $v \in \V$ define an assignment $\phi(v)$ to be the set of signals returned from sensor $v$ for variables in $\D(v)$.  We wish to study the consistency of $\phi$ over the network. For each $\sigma \in \X$ with vertex set $V' \subset \V$ we compare the values 
$$\{\shr{\str{\sigma} \subset \text{star}(v)}(\phi(v))\}_{v \in \sigma}.$$

To illustrate, suppose we have two weather stations, represented as $v_1$ and $v_2$, each returning meteorological measurements. Suppose $v_1$ returns temperature and atmospheric pressure while $v_2$ returns temperature and wind speed. We represent this assignment as $\phi(v_1) = (t_1,p_1)$ and $\phi(v_2) = (t_2,w_2)$. Because the sensors both return temperature readings we represent the relationship between $v_1$ and $v_2$ with a hyperedge $\sigma$. To determine if the weather stations are returning consistent data we might compare
$$
t_1 = \shr{\str{\sigma} \subset \text{star}(v_1)}(\phi(v_1))  \stackrel{?}{=}
\shr{\str{\sigma} \subset \text{star}(v_2)}(\phi(v_2)) = t_2.
$$

But what do we mean by consistency? Two temperatures may be considered consistent if they are equal, or if they are within some $\epsilon$ of each other. The data need not even be numeric. One weather station may simply return a categorical rating indicating if the temperature is warm or cold.  Without loss of generality we assume that the data being returned is preprocessed into common units and categories. Beyond this we will assume nothing about the definition of consistency. As in \cite{Robinson2} we suppose that someone has decided what will be considered consistent when comparing the assignment to variables of vertices belonging to the same cell. This someone will provide a rule $C_{\sigma}$ for each cell $\sigma \in \X$ which returns a boolean value, $0$ for inconsistent and $1$ for consistent. 

\begin{defn*}
A \textit{consistency structure} $(\X,\shf,\C)$ is a cell complex $\X$, a sheaf $\shf$ over $\X$, and a set of consistency functions $\C = \{\C_{\sigma}\}_{\sigma \in \X}$, which 
return boolean values on the $k$-cells in $\X$ for $k \geq 1$. That is for each cell $\sigma$ defined by a set of vertices $\V_i$ we have: 
$$\C_{\sigma} : \bigoplus_{v \in \V_i} \shm{\str{\sigma} \subset \str{v}}(\shm{\str{v}}) \longrightarrow \{0,1\}.$$
\end{defn*}

Let $\phi$ define an assignment of data to the vertices in $\V$. We evaluate the consistency of $\phi$ on each cell $\sigma$ by computing $$\C_{\sigma}(\phi) = \C_{\sigma}(\shm{\str{\sigma} \subset \str{v_0}}(\phi(v_0)),...,\shm{\str{\sigma} \subset \str{v_k}}(\phi(v_k))).$$ We say that $\phi$ is a consistent assignment on $\sigma$ if and only if $\C_{\sigma}(\phi) = 1$.

The \textit{standard consistency structure} $(\X,\shf,\C^s)$ is defined on an assignment $\phi$ so that $\C^s_{\sigma}(\phi) = 1$ if and only if there exist a unique section $s_{\sigma}^{\phi} \in \shm{\str{\sigma}}$ such that $\shm{\str{\sigma} \subset \text{star}(v)}(\phi(v)) = s_{\sigma}^{\phi}$ for all vertices $v \in \sigma$. This is the case where by consistent we really mean exactly the same. In the weather station example, 
$\C^s_{\sigma} = 1$ only when $t_1 = t_2$ but there could be a different consistency structure for which $\C_{\sigma} = 1$ if the temperatures are deemed to be \emph{close}. 

For any $\V_i \subset \V$, let $\Y_i$ be the subcomplex of $\X$ induced by the vertex set $\V_i$. In particular, $\Y_i$ consists of all of the cells in $\X$ which are defined by subsets of $\V_i$. $\Y_i$ is an abstract simplicial complex which inherits a subspace topology from $\X$ and a consistency structure $(\Y_i, \shm{\Y}, \C|_{\Y_i})$, where $\C|_{\Y_i}$ is the set of boolean functions belonging to $\C$ defined on the cells of $\Y_i$. Let $\phi$ be an assignment on $\V$. We say $\phi$ is \textit{consistent on $\Y_i$} if $\phi$ is a consistent assignment on each cell in $\Y_i$. We would like to identify the largest subcomplexes in $\X$ on which $\phi$ is consistent.

\begin{thm*}

Let $(\X,\shf, \C)$ be a consistency structure for a finite abstract simplicial complex $\X$ with vertex set $\V$. Let $\phi$ be an assignment on $\X$. Then there exists a unique collection of subsets $\{\V_{\alpha}\}$ of $\V$ which induce subcomplexes $\{\Y_{\alpha}\}$ of $\X$ with the following properties.
 \begin{enumerate}
 \item The assignment $\phi$ is consistent on each $Y_{\alpha}$, and any subcomplex of $\X$ on which $\phi$ is consistent is a subcomplex of one of the $Y_{\alpha}$.
 \item The open sets $\{\str{Y_{\alpha}}\}$ form a cover of $\X$.
 \item If $(\X,\shf, \C^s)$ is a standard consistency structure, then for each $\Y_{\alpha}$ there is a unique local section $s_{\alpha}^{\phi} \in \shm{\str{Y_{\alpha}}}$ such that 
 $\shm{\str{v} \subset \str{Y_{\alpha}}}(s_{\alpha}^{\phi}) = \phi(v)$, for each vertex $v$ in $\V_{\alpha}$.
 \end{enumerate}

\end{thm*}

The proof is constructive. There are a finite number of cells in $\X$. We use $\C$ to identify cells on which $\phi$ fails to give a consistent assignment. We construct subcomplexes of $\X$ which do not contain these cells but are careful not to break up any subsets of the vertices on which $\phi$ does give a consistent assignment. Since $\X$ has a finite number of cells, this process must terminate. The basic operation is described in the following lemma.

\begin{lem*}

Let $(\Y,\shf, \C)$ be a consistency structure for a finite abstract simplicial complex $\Y$ with vertex set $\V$. Let $\phi$ be an assignment on $\Y$. Suppose $\sigma$ is a cell in $\Y$ generated by vertices $\V_i \subset \V$ such that $\C_{\sigma}(\phi) = 0$. Then there exist subsets $\{\V_{\alpha}\}$ of $\V$ with induced subcomplexes $\{\Y_{\alpha}\}$ of $\Y$ such that:

\begin{enumerate}
\item $\sigma$ does not belong to any $\Y_{\alpha}$.
\item Every subcomplex on which $\phi$ gives a consistent assignment is a subcomplex of at least one $\Y_{\alpha}$. 
\item The set $\{\str{\Y_{\alpha}}\}$ is a cover of $\Y$.
\end{enumerate}

\end{lem*}
\noindent
\emph{Proof: }
Suppose $\sigma$ has vertex set $\V_i = \{v_{0}, v_{1},...,v_{k}\}$. Since $\C_{\sigma}(\phi) = 0$ no consistent subcomplex of $\Y$ may contain $\sigma$ as a cell. For each vertex $v_i$ in $\sigma$ let $\V_{i} = \V \setminus \{v_{i}\}$ and $\{\Y_{i}\}$ be the corresponding set of induced subcomplexes of $\Y$. Clearly $\sigma$ does not belong to any of the $\Y_{i}$, and if $\Y'$ is a subcomplex of $\Y$ on which $\phi$ is consistent then the vertices in $\Y'$ must be a subset of one of the $\V_{i}$. 
Since $\Y = \cup_{v \in \V}\str{v}$ and every vertex in $\V$ is contained in at least one of the $\Y_{i}$ we have $\Y = \cup \str{\Y_{i}}$. This proves the lemma.

\vspace{5mm}
\noindent
\emph{Proof of Theorem: }
We prove the theorem by induction, repeatedly applying the lemma to $\X$ and its subcomplexes to form a list of subcomplexes which have the properties returned by the lemma and noting that the process must stop because $\X$ contains a finite number of cells. 

Suppose $C_{\sigma}(\phi) = 0$ for some cell $\sigma$ in $\X$ and $\sigma$ is a $k$-cell. We construct a list of $k$ subsets $\{V_{\alpha}\}$ of $\V$ and a list of the corresponding $k$ subcomplexes $\{\Y_{\alpha}\}$ which satisfy the lemma. Within each $\Y_{\alpha}$ we check for consistency by computing $\C_{\tau}(\phi)$ for each cell $\tau$ in $\Y_{\alpha}$. If we find one case where $\C_{\tau}(\phi) = 0$ then we apply the lemma and replace $\V_{\alpha}$ in the list of vertex sets with its subsets returned by the lemma and $\Y_{\alpha}$ in the list of subcomplexes with its subcomplexes returned by the lemma. 

By design, the relative complements of any pair of the original $\{V_{\alpha}\}$ consist of a single vertex. Their pairwise intersections contain the balance of the vertices, hence their induced subcomplexes $\{\Y_{\alpha}\}$ will share cells.  Since we are only interested in maximal consistent assignments we drop any subcomplex from the list whose vertices form a proper subset of the vertices of another subcomplex in the list.

We repeat this process on the new list and continue until no subcomplex contains a cell $\tau$ for which $\C_{\tau}(\phi) = 0$. Let $\{\V_{\beta}\}$ be the list of subsets of $\V$ and $\{\Y_{\beta}\}$ be the list of subcomplexes of $\X$ obtained in this way. By construction $\phi$ induces a consistent assignment on each $\Y_{\beta}$. By the lemma, any subcomplex on which $\phi$ induces a consistent assignment belongs to one of the $\Y_{\beta}$ and the set $\{\str{\Y_{\beta}}\}$ is a cover of $\X$. If $(\X,\shf, \C)$ is a standard consistency structure then we apply the gluing property of $\shf$ to uniquely define $s_{\beta}^{\phi} \in \shm{\str{Y_{\beta}}}$ so that $\shm{\str{v} \subset \str{\Y_{\beta}}}(s_{\beta}^{\phi}) = \phi(v)$, for each $v$ in $\V_{\beta}$. 

To show uniqueness we suppose  $\{\V_{\gamma}\}$ is a collection of subsets of $\V$ such that the set of corresponding induced subcomplexes $\{\Y_{\gamma}\}$ satisfy the properties of the theorem. Since $\phi$ is consistent on each $\Y_{\beta}$ there is a $\Y_{\gamma}$ such that $\Y_{\beta}$ is a subcomplex of $\Y_{\gamma}$. Since $\phi$ is consistent on $\Y_{\gamma}$ there is a $\hat{\Y_{\beta}}$ such that $\Y_{\gamma}$ is a subcomplex of $\hat{\Y_{\beta}}$. Consider the corresponding sets of vertices. We have
$$\V_{\beta} \subset \V_{\gamma}  \subset \hat{\V_{\beta}}.$$
But the list of $\{\V_{\beta}\}$ contains no proper subsets by construction. Hence $\V_{\beta} = \V_{\gamma}  = \hat{\V_{\beta}}$ and $\Y_{\gamma} = \Y_{\beta}$. This proves the theorem.

\vspace{5mm}
The theorem provides a systematic way to split the cell complex into a unique set of maximal subcomplexes of $\X$ on which a single assignment of data is consistent. 

\begin{figure}[H]
\centering
\includegraphics[scale=.5]{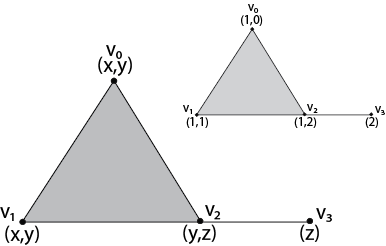}
\caption{For examples 1 and 2, this figure depicts a sensor network with four sensors returning numeric data values for variables $x$, $y$, and $z$. Cells exist where sensors return values for the same variables. }
\label{ex1}
\end{figure}

\begin{ex}\label{ex01}
Let $\X$ be the abstract simplicial complex pictured in Figure \ref{ex1}. The complex $\X$ has four $1$-cells and one $2$-cell. Let $\shf$ be the sheaf of data assignments over $\X$ and $(\X,\shf,C^s)$ be the standard consistency structure on $\X,\shf$. 

Let $\phi$ be an assignment on the vertices given by: $\phi(v_0) = (x=1,y=0)$, $\phi(v_1) = (x=1,y=1)$, $\phi(v_2) = (y=1,z=2)$, and $\phi(v_3) = (z=2)$, as labeled in the smaller copy of the complex in Figure \ref{ex1}.  We find $C^s_{v_0 v_1}(\phi) = 0$, $C^s_{v_1 v_2}(\phi) = 1$, $C^s_{v_0 v_2}(\phi) = 0$, $C^s_{v_2 v_3}(\phi) = 1$, and $C^s_{v_0 v_1 v_2}(\phi) = 0$. The maximal vertex sets which induce subcomplexes on which $\phi$ is consistent are: $\V_1 = \{v_0,v_3\}$ and $\V_2 = \{v_1,v_2,v_3\}$.
\end{ex}

\begin{ex}
Let $\X$, $\shf$, $\phi$ be defined as in Example \ref{ex01}. Define a consistency function $\C$ on each cell in $\X$ by 
$C_{\sigma} = 1$ if and only if the data values returned by the vertices generating $\sigma$ agree on at least one variable. In this case  
$C_{v_0 v_1}(\phi) = 1$, $C_{v_1 v_2}(\phi) = 1$, $C_{v_0 v_2}(\phi) = 0$, $C_{v_2 v_3}(\phi) = 1$, and $C_{v_0 v_1 v_2}(\phi) = 0$. The maximal vertex sets which induce subcomplexes on which $\phi$ is consistent in this example are: $\V_1 = \{v_0,v_1,v_3\}$ and $\V_2 = \{v_1,v_2,v_3\}$.
\end{ex}

\begin{figure}[H]
\centering
\includegraphics[scale=.5]{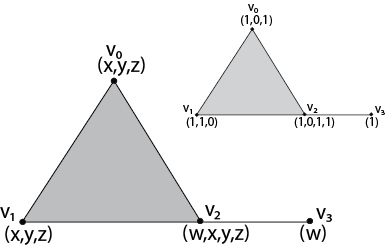}
\caption{For examples 3 and 4, this figure depicts another sensor network with four sensors returning numeric data values for variables $w$, $x$, $y$, and $z$. }
\label{ex2}
\end{figure}

\begin{ex}\label{ex02}
Let $\X$ be the abstract simplicial complex pictured in Figure \ref{ex2}. The complex $\X$ has four $1$-cells and one $2$-cell. Let $\shf$ be the sheaf of data assignments over $\X$ and $(\X,\shf,C^s)$ be the standard consistency structure on $\X,\shf$. 
Let $\phi$ be the assignment given in the smaller copy of the complex in the figure. Clearly the only $k$-cell ($k\geq1$) on which $\phi$ is consistent is $\{v_2,v_3\}$. Hence the maximal vertex sets which induce subcomplexes on which $\phi$ is consistent in this example are: $\V_1 = \{v_0,v_3\}$, $\V_2 = \{v_1,v_3\}$ and $\V_3 = \{v_2,v_3\}$. 
\end{ex}

\begin{ex}\label{ex03}
Let $\X$, $\shf$, $\phi$ be defined as in Example \ref{ex02}. Define a consistency function $\C$ on each cell in $\X$ by 
$C_{\sigma} = 1$ if and only if the data values returned by the vertices generating $\sigma$ agree on at least one variable. In this case the only cell on which $\phi$ is not consistent is $\sigma = \{v_0,v_1,v_2\}$. This means that even though $\phi$ is consistent on each of the $1$-cells in $\X$, the $1$-cells belonging to the boundary of $\sigma$ cannot belong to the same subcomplex. The resulting maximal vertex sets which induce subcomplexes on which $\phi$ is consistent in this example are: $\V_1 = \{v_0,v_1,v_3\}$, $\V_2 = \{v_0,v_2,v_3\}$ and $\V_3 = \{v_1,v_2,v_3\}$. 

\end{ex}

\begin{figure}[H]
\centering
\includegraphics[scale=.5]{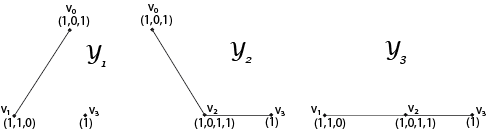}
\caption{The maximal consistent subcomplexes of the complex $\X$ under the assignment $\phi$ in Example \ref{ex03} }.
\label{ex3}
\end{figure}

\bibliographystyle{plain}
\bibliography{thispaper}
\end{document}